\documentclass{article}
\usepackage{graphicx} 

\title{Agency cannot be a purely quantum phenomenon}

\date{\today}
\usepackage{natbib}
\usepackage{amsmath}
\usepackage{amsfonts}
\usepackage{braket}
 \usepackage{physics}
\usepackage{subcaption}
\usepackage{authblk}

   \usepackage[top=2.5cm, bottom=2.5cm, left=2.5cm, right=2.5cm]{geometry}
 \usepackage{setspace}
\begin{document}


\author[1,2]{Emily C. Adlam}
\author[1,2]{Kelvin J. McQueen}
\author[1]{Mordecai Waegell}

\affil[1]{\small Institute for Quantum Studies, Chapman University}
\affil[2]{\small Philosophy Department, Chapman University}

\maketitle

\begin{abstract}
What are the physical requirements for agency? We investigate whether a purely quantum system (one evolving unitarily in a coherent regime without decoherence or collapse) can satisfy three minimal conditions for agency: an agent must be able to create a world-model, use it to evaluate the likely consequences of alternative actions, and reliably perform the action that maximizes expected utility. We show that the first two conditions conflict with the no-cloning theorem, which forbids copying unknown quantum states: world-model construction requires copying information from the environment, and deliberation requires copying the world-model to assess multiple actions. Approximate cloning strategies do not permit sufficient fidelity or generality for agency to be viable in purely quantum systems. The third agency condition also fails due to the linearity of quantum dynamics. These results imply four key consequences. First, agency requires significant classical resources, placing clear constraints on its physical basis. Second, they provide insight into how classical agents emerge within a quantum universe. Third, they show that quantum computers cannot straightforwardly simulate agential behavior without significant classical components. Finally, they challenge quantum theories of agency, free will, and consciousness.  
\end{abstract}

\tableofcontents

\newpage

\section{Introduction}

What are the physical requirements for agency? In this paper we examine whether a purely quantum agent is possible. By this we mean an agent realized in a purely quantum system: one that evolves unitarily in a coherent regime without decoherence or collapse. Much of our analysis will assume that the agent's environment is also purely quantum, though later we relax this assumption.

For a purely quantum system operating in a purely quantum environment, no preferred basis is available either within the system or in the environment, so the agent cannot take advantage of  environmental states occurring in some specific orthogonal basis. This is precisely the regime in which results such as the no-cloning theorem apply. In real physical systems, preferred bases often emerge through decoherence: states converge toward stable bases either under persistent monitoring by an external environment or through the effective self-decoherence of large, complex systems with many internal degrees of freedom. Small coherent systems, by contrast, do not exhibit such convergence. We therefore focus on the regime without an emergent preferred basis, which we regard as a hallmark of classicality. 

Could a purely quantum system be an agent? We argue that it cannot. We characterize agency in terms of three minimal conditions. First, an agent must be able to construct a world-model of its environment. Second, it must be able to use that model to evaluate the likely consequences of alternative actions. Third, it must be able to reliably select and implement the action that maximizes expected utility. These conditions are deliberately modest, so that they apply to both artificial and natural agents, and they align with insights from philosophy, normative decision theory, artificial intelligence, as well as empirical work in neuroscience and cognitive science. Our aim is not to insist that this is the only concept of agency, but to examine whether even this minimal and significant form of agency is possible in a purely quantum system.

We show that the first two agency conditions conflict with the no-cloning theorem, which forbids copying arbitrary unknown quantum states. World-model construction requires copying information from the environment, and deliberation requires copying the world-model to test different possible actions. We examine several strategies for avoiding this conclusion. One is to suppose the agent extracts environmental states wholesale, as in quantum swap operations. But once extracted, these states are no longer part of the environment, and so cannot be relied upon to guide action. Another strategy is to posit that the agent has access to multiple identical copies of the relevant quantum state, so that it need not make copies itself. This would allow the agent to simulate multiple actions, but it does not help with the third agency condition: the linearity of quantum dynamics prevents any unitary operation from comparing the simulations and reliably implementing the `best' action. We also explore approximate cloning schemes. These approaches provide only limited workarounds for the first two conditions, but none offer the generality or fidelity needed to sustain reliable agential behavior.

More broadly, our results have four key consequences. First, they place principled constraints on the physical possibilities for agency. We do not yet have a complete theory of mind or agency, but we can make progress by ruling out regimes where agency cannot be realized; our results cut down the space of options by showing that agency cannot be sustained in a purely quantum regime. Second, they help to clarify how agents could emerge within a quantum universe: agency requires classical features such as a preferred basis that can support the reliable copying of information, so the emergence of such classical structures marks the point at which agential behavior becomes physically possible. Third, they establish a technological limitation. Quantum computers are built with respect to a computational basis imposed by classical control, but this basis does not by itself ensure that environmental states naturally occur in that basis; such alignment requires decoherence. In practice, quantum algorithms rely on classical users to prepare inputs in the computational basis and to interpret outputs accordingly. Thus, a quantum computer that simulates agency would have to rely heavily on classical external control to provide inputs in a preferred basis, and it would not be effective if tasked with sampling at random from a quantum environment with no preferred basis. Finally, our results challenge quantum theories of agency, free will, and consciousness, forcing such proposals to identify explicitly where the necessary classical resources enter into their account, and to explain why the quantum resources they invoke are not thereby rendered redundant. 

The paper proceeds as follows. Section 2 clarifies the conception of agency at issue. Section 2.1 sets out our three conditions for agency and situates them within decision theory. Section 2.2 contrasts this view with looser conceptions of agency, and explains why our more demanding notion is both significant and well-motivated. Section 3 turns to the physical constraints on purely quantum agents. Section 3.1 reviews the no-cloning theorem and its implications for copying quantum states. Section 3.2 explains our notion of a purely quantum system in more detail and then outlines the challenges that the no-cloning theorem poses for world-model construction and deliberation in such a system. Section 3.3 considers the case of agents with access to unlimited perfect copies, and proves that even then the third agency condition fails due to the linearity of quantum dynamics. Section 3.4 explores approximate and probabilistic cloning strategies. Section 3.5 evaluates some proposed quantum agency circuits and their performance. Section 4 draws out the broader implications of our results, focusing especially on how they challenge quantum theories of agency, free will, and consciousness.

\section{Agency}

\subsection{The concept of agency}

In this paper we will assume three necessary conditions on agency: an agent must be able to create a world-model, use it to evaluate the likely consequences of alternative actions, and reliably perform the action that maximizes expected utility. A world-model is an internal representation of the external world, including one's own place in it. Agents use world-models to assess the likely consequences of their actions and thus to decide which action to take. 

For example, suppose I am trying to decide whether to walk to campus or take the bus. My decision-making is going to involve first gathering information - what does the weather forecast say, is the bus running on time - and making a simple mental model of what might happen if I take each of these two options, in order to determine the likely outcomes. If I walk, I'll get some exercise, but there's 
a decent possibility that it will rain and my books will get wet. If I take the bus, I don't have to worry about the weather, but there's a decent possibility that the bus will not be on time and I'll be late. I can then weigh up how desirable or undesirable I find each of the possible outcomes and make a decision about what to do. 

It is possible to formalize this decision-making process using the standard framework of decision-theory \citep{Savage1954, Peterson2017}. An agent takes in sensory inputs and uses them to create a representation of the state of her immediate environment, as well as relevant parts of the wider world, and her own position in it. The agent then creates copies $\{ C_i \}$ of this representation and for each copy $C_i$, she models herself performing a possible action $A_i$; she can then assign probabilities $\{ p^i_j \} $ to various different outcomes $O_j$ that might follow from this action. For example, if she is confident that action $A_i$ will definitely produce a certain result $O_j$, she will assign $p^i_j = 1$ and $p^i_k = 0$ for all $k \neq j$. 

Let us assume that the agent has a fixed set of values or desires. These provide a measure of how much she prefers or finds undesirable each possible outcome, which we represent as utilities $U^i_j$ assigned to outcome $O_j$ conditional on action $A_i$. Thus the agent can calculate the expected utility $E_i$ of every possible action: 

\[ E_i = \sum_j p^i_j U^i_j \]

The agent can then simply perform the action which leads to the highest expected utility $E_i$. As a simple illustration, if I believe there is a 0.8 chance of rain, then I will conclude that there is a 0.8 chance of getting wet if I choose to walk. Getting wet has a strong negative utility (say, --10), while getting exercise has a positive one (say, +5). Assuming I only care about staying dry and getting exercise, the total utility of walking if it rains is --5, and if it does not rain it’s +5. The expected utility of walking is therefore $(0.8\times-5)+(0.2\times5)=-3$. If the expected utility of taking the bus is higher, then the rational choice is to take the bus.

It is well-known that this kind of decision-theoretic story is an oversimplification. First of all, in most circumstances agents do not perform precise quantitative calculations, so the formula given above should be thought of as an (unrealistic) idealization rather than a literal story about the exact thought processes of a realistic agent. In addition, there is abundant evidence that real agents are not perfectly rational and thus sometimes they may fail to perform the action which leads to the highest expected utility, or they may ascribe utility to outcomes in an unexpected or irrational way. However, it nonetheless seems fair to say that the decision-theoretic model captures at least the approximate form of many important decision-making processes. And our third agency condition only requires that the agent has the ability to reliably perform the action the maximizes expected utility, not that they are guaranteed to do so on every occasion. 

For our purposes we do not need to use the full framework of decision theory. Perhaps our most important assumption is that to deliberate, the agent creates copies of their world-model: for each copy $C_i$, the agent models itself performing a possible action $A_i$. This idea of ``copying" may sound trivial, because in a classical setting copying is effortless—we can freely take stored data, reproduce it in another register or location, and then manipulate the copy while leaving the original intact. But precisely this ease hides an important fact: deliberation requires branching the same information into multiple scenarios, and this branching presupposes the ability to duplicate representations. Later (in section 3.1) we will see that this assumption, so natural in a classical framework, cannot simply be carried over to the quantum case, since there the act of copying information is no longer straightforwardly available.

In what follows, we will make one further simplification. We set aside the subjective probabilities that feature in standard decision theory, and instead assume that for any given action and world-model the agent is always certain about what the outcome will be. This is because our agents are assumed to be in a deterministic universe, with complete knowledge of the consequences of their possible actions, given a specification of the state on which they are acting. This allows us to focus on the structure of deliberation itself without the added complexity of probabilistic reasoning. At the same time, this simplification actually makes things easier for a putative quantum agent, since probabilities would require even more copying: to represent the possibility of rain when walking to campus, for example, the agent would need to model both the ``rain" scenario and the ``no-rain" scenario separately. Thus introducing probabilities increases the number of scenarios that must be considered, and hence multiplies the copying requirement.\footnote{It has also been found that copying occurs even once a decision has been formed: motor commands are accompanied by an efference copy (or corollary discharge) transmitted to other brain regions, such as the hippocampus, where it provides an internal record of the intended action and its predicted consequences \citep{numan2015prefrontal}. Such mechanisms are thought to underlie our ability to monitor and distinguish self-generated from external events, and disruptions of efference copying have been linked to the disturbed sense of agency in schizophrenia \citep{pynn2013function}.}

A separate issue concerns whether the deliberative process itself could be fundamentally deterministic or indeterministic. The environment might itself contain genuinely indeterministic events, or the agent’s decision procedure might incorporate stochastic elements—for example, instead of always selecting the maximum-utility option, the agent might sometimes select actions probabilistically in proportion to their expected utilities. Our framework is neutral on this matter: the models may be deterministic or indeterministic, and the agent may select deterministically or with some randomness. The only restriction is that the modelling and selection cannot be entirely random if they are to be useful to the agent, since some structure is required for goal-directed behavior \citep{Durham_2020}. Thus our conception of agency is compatible with both determinism and indeterminism, and does not presuppose any particular theory of free will (e.g. libertarianism or compatibilism).

\subsection{Comparison with other concepts of agency}

The concept of agency is employed across many disciplines, from philosophy and psychology to biology and artificial intelligence, and unsurprisingly the term is used in different ways. Our aim in this section is to situate our own conception within this broader landscape. We do not claim that there is a single correct notion of agency and we allow that different fields may require different concepts for their purposes. But we will argue that the conception we adopt is both significant and well-motivated. In particular, it captures the model-based deliberation that many disciplines—philosophy, psychology, neuroscience, and AI—identify as a hallmark of genuine agency.

Some approaches in biology extend the notion of agency to surprisingly minimal or even non-intuitive cases. For example, \citet{levin2024self} has argued that even memories (or engrams) can count as agents. As an example, a traumatic memory can preserve and propagate itself in the brain despite the individual’s attempts to suppress it. The memory maintains itself homeostatically, in the way a simple organism might maintain its integrity against external disruption. This illustrates how some definitions treat agency as equivalent to self-maintenance. While such a broadened notion can be illuminating, especially in biological contexts, it is far removed from the richer, deliberative conception that is our focus.

Some approaches in artificial intelligence extend the notion of agency to systems as simple as thermostats. For example, \citet{RussellNorvig2003} provide a useful classification of agents into five kinds arranged by increasing sophistication. The most basic are \textit{simple reflex agents}, which act only on the basis of their immediate percepts, using fixed condition–action rules. A thermostat is a standard example: if the temperature drops below a set point, it turns on the heat. A robot vacuum that turns and continues whenever it encounters an obstacle also fits this category. The next step up are \textit{model-based reflex agents}, which supplement condition–action rules with a simple internal model. For example, a robot vacuum might keep track of where it has already been, so that it does not repeatedly vacuum the same spot. More sophisticated still are \textit{goal-based agents}, which use a model of the environment to evaluate whether potential actions achieve some specified aim—for instance, a robot taxi that compares alternative routes through a city until it finds one that reaches the passenger’s destination by a certain time. Finally, \textit{utility-based agents} extend this approach by assigning utilities to possible outcomes and selecting the action with the highest expected utility: the robot taxi might assign scores to different routes based not just on whether they reach the destination, but also on their expected time, cost, and comfort, and then choose the route with the best overall trade-off. This last category most closely aligns with the conception of agency we are defending here. (Russell and Norvig also identify a fifth category, the learning agent, but we do not focus on learning in this paper.)

With all of these different notions of agency available, why do we focus on the utility-based notion that requires agents to satisfy our three conditions? The answer is that our aim is not to explain what makes a thermostat, a cell, or even a memory ``agent-like" in some broad and metaphorical sense. Our concern is with the kind of agency that we ourselves manifest in everyday deliberation: the capacity to model the world, compare the consequences of alternative actions, and then select the option that best advances our goals. This is the form of agency that underlies decision-making in humans and sophisticated artificial systems, and it is precisely this form of agency that quantum theories of mind and free will aim to capture. By focusing on this conception, we put ourselves in a position to ask whether the basic structure of our own agency could, in principle, be realized in a purely quantum system. That is why this notion is not only significant, but indispensable for our purposes.

Support for our conception of agency also comes from multiple scientific and philosophical domains. In philosophy, naturalistic accounts of free will and moral responsibility typically presuppose the kind of deliberative agency we have described. For example, \citet{Frankfurt1969} emphasizes the importance of higher-order reflection on one's motives, while \citet{Fischer1998-FISRAC-3} develop a reasons-responsive account of agency, according to which responsible agents must be capable of recognizing and responding to reasons that bear on their actions. Similarly, \citet{kane2007libertarianism} highlights the role of evaluating alternatives in his idea of ``self-forming actions". Though these accounts differ in detail, they converge on the idea that agency centrally involves the capacity to represent possible courses of action, assess their consequences, and select accordingly.

Psychology and neuroscience provide complementary empirical support \citep{beni2025agency}. Experimental work on the ``sense of agency" uses intentional binding (where causes and their effects are perceived as closer in time when the subject takes herself to be the cause) as an objective measure of agential experience. \citet{kulakova2017could} show that participants’ sense of agency is strongest when they have alternative actions leading to distinct possible outcomes, suggesting that deliberative choice enhances agential experience. Neuroscience reinforces this connection: \citet{haggard2017sense} finds evidence that the dorsolateral prefrontal cortex, a region associated with selecting among competing actions, is closely linked to the subjective sense of agency. Taken together, these results indicate that the mechanisms underpinning our sense of agency are not triggered merely by condition–action rules or self-maintenance, but specifically by processes that involve representing alternatives and comparing their outcomes.

Recent results in AI theory also support our conception of agency. \citet{richens2025generalagentsneedworld} prove that any system capable of performing long-horizon, goal-directed tasks in complex environments must in effect possess an accurate predictive model of its environment, regardless of how it is trained or implemented. Meanwhile, \citet{richens2024robust} establish a complementary result: an agent capable of adapting to a sufficiently wide range of distributional shifts---that is, generalizing across environments---must have learned a causal world-model. Taken together, these proofs show that both task generalization and domain generalization entail world-modeling.

Now, one might think that an agent can in effect behave as if it has learned a causal world-model without continuously modelling on the fly: it could simply go to a lookup table in its internal storage which directly specifies an action for each input, and output the specified action. But as a matter of fact, this will not work for realistic agents in complex environments. Suppose there is a function $f$ that, given a description of the current state of the environment $I$, always selects the action that maximizes some outcome measure $Q$. Here $Q$ can be thought of as overall utility: in the walk versus bus case, $Q$ combines factors such as whether my books get wet, whether I get exercise, and whether I arrive on time. For $f$ to succeed systematically across many possible environments, it cannot merely store a brute mapping from each possible $I$ to the best action. Such a lookup table would be infeasible in any complex setting, since it would need to contain an entry for every possible variation in weather, bus reliability, timing, and goals. Instead, $f$ must encode how the value of $Q$ changes under different actions given different features of $I$. For example, it must represent that ``rain decreases the utility of walking'' and that ``bus delays decrease the utility of taking the bus.'' Formally, this means that for every input $I$ and every possible action $A_x$, $f$ presupposes access to a mapping $g_I(A_x) \mapsto Q$, where $g_I$ represents \textit{how the environment would evolve} under action $A_x$. In other words, $f$ must already contain a structure that predicts the outcome of each action. Thus, the ability to select the best action across a wide range of circumstances is only possible if the system carries, in effect, an internal world-model and performs modelling continuously on the fly. In terms of resources, a lookup table of memorized contingencies is only practical/plausible for relatively simple environments, and thus dealing with richly complex environments requires world-modeling.

In particular, the lookup table approach gives systems quite a limited ability to respond dynamically to inputs or goals they have not seen before. When faced with unfamiliar data or when asked to achieve a new type of goal, a system using a lookup table approach can do nothing other than apply some kind of naive interpolation. In cases where the environment has the kind of structure which is friendly to interpolation this method may sometimes produce reasonable results, but it has obvious limitations — for example, features of the environment may sometimes interact in non-obvious ways which leads to outcomes quite different from what one might expect based on a naive interpolation. Clearly the success of this kind of naive interpolation depends strongly on the complexity of the environment which is generating the rewards — for example if the function is linear but the environment exhibits strong non-linearities, naive interpolations are unlikely to be very successful. 

By contrast, a system which has the capacity to directly model its  complex environment and make new predictions on the fly for the impacts of various kinds of actions will generally have a much better ability to determine the right way to respond to new kinds of input or to achieve new types of goals, since it can take into account environmental interactions that it has not previously experienced directly, and it can appeal to principled reasons to determine which inputs are relevantly similar. So the latter system is much more like what we would normally think of as an agent, in that it can make reasoned decisions about what to do in novel situations or with novel goals, rather than just doing naive interpolation. 

We can also see the relevance of this distinction by looking at machine-learning models. We can do a simple kind of `machine-learning' by fitting a curve to some data: we then have a fixed function from inputs to outputs, and when given a new input, the system just applies this function to generate an output. Suppose instead that we put the data into a neural network. If the neural network has a simple structure or it is not trained for a long time, it might learn something which is essentially just simple curve fitting. But if our neural network has a large enough network of weights and is trained on sufficiently complex data, its weights are likely to develop some kind of internal structure which can be thought of as making multiple copies of the data and modeling the impact of various different choices. For if the inputs and outputs are produced by some kind of structured environment, the results of \citet{richens2025generalagentsneedworld} indicate that the network must in effect be employing a causal world-model of that environment. 

Note in particular that although a neural network does not explicitly make copies of its input data, in practice it requires a lot of `copying' of the kind at issue here. This is because in a general neural network, a given node $n_o$ on one layer of the neural network may be connected to many nodes $n_1, n_2 ... n_n$ on the next layer, meaning that the values on the nodes  $n_1, n_2 ... n_n$ depends on the value of $n_o$. Calculating the value of $n_1$ from $n_o$ is a kind of information-processing, so in effect we require a `copy' of $n_o$ for each of the nodes $n_1, n_2 ... n_n$.

Now, in a classical context one might be inclined to say that the line between these two cases is not perfectly well-defined. After all, even when a system does model its environment it is to a large degree relying on looking for similarities to previous cases, albeit in a more complex way than a function which simply maps inputs to outputs. In this sense, the distinction between agential/non-agential as described above comes in degrees. 

However, note that the distinction becomes particularly important in the context of a potential quantum agent in a quantum environment. For if a quantum agent receives an input in the form of an unknown quantum state, the agent cannot, as we will now argue, make perfect copies of that state, and so its ability to perform multiple calculations and compare the results is very limited.  So there is in fact a physically meaningful and sharp distinction between the two cases: the single-calculation case does not require copying and thus can be implemented exactly in a purely quantum setting, whereas the multiple-calculation case requires copying and thus can only be imperfectly implemented in a purely quantum setting.

\section{Quantum agency}

\subsection{The no-cloning theorem}

The concept of `copying' is central to our concept of agency. To illustrate it, imagine that I have a classical variable stored in a register, and I consult that register multiple times in order to perform various kinds of information-processing. For example, perhaps this variable encodes the population of California and I use it once to calculate the estimated annual energy consumption of California, and then I use it again to calculate the estimated annual banana consumption of California, and then a third time to calculate the estimated annual greenhouse gas emissions from California. Classically, we might not think of this process as involving `copying' - we are simply consulting the register multiple times. 

But in fact the process does in an important sense involve copying: for each calculation I made a copy of the classical variable and then performed various mathematical transformations on it to produce a desired output. In a classical context copying is unlimited and free so we don't usually think about this process explicitly in terms of copying data. But we can see that it involves copying if we imagine a comparable situation with an unknown quantum state stored in a register. If I want to use that state to do information processing I must take it out of the register and perform operations on it to produce my desired final result, and once I have done so it has been used up: I cannot perform a different calculation with it unless I simply undo the operations that I have performed, and if I do, then I have erased the outcome of my calculation. 

The no-cloning theorem places a strong constraint on the possibility of copying a quantum state $\psi$, a procedure we can write as $\psi \rightarrow \psi \otimes \psi$. The theorem can be understood as stating that: \textit{cloning can be done perfectly and with probability 1 if and only if a basis to which $\psi$ belongs is known} \citep{Scarani_2005}. 

The theorem can also be stated more generally: \textit{no quantum operation exists that can perfectly copy an arbitrary quantum state}. The proof of this theorem in terms of the linearity of quantum operations is straightforward \citep{wootters1982single}. Consider a machine which clones the two orthogonal states $\ket{1}$ or $\ket{0}$, by copying the original state into a ``ready" state: $\ket{0}\ket{R} \rightarrow \ket{0}\ket{0}$, $\ket{1}\ket{R} \rightarrow \ket{1}\ket{1}$. We can now ask what the machine does to a superposition of these states, $\ket{\psi} = \alpha\ket{0} + \beta\ket{1}$? By linearity, we have:

\[ \ket{\psi}\ket{R} \rightarrow \alpha\ket{0}\ket{0} + \beta\ket{1}\ket{1}\]

But, on the other hand:

\begin{align}
\ket{\psi}\ket{\psi} 
  &= (\alpha\ket{0} + \beta\ket{1})(\alpha\ket{0} + \beta\ket{1}) \nonumber \\
  &= \alpha^2\ket{0}\ket{0} + \alpha\beta\ket{0}\ket{1} 
     + \beta\alpha\ket{1}\ket{0} + \beta^2\ket{1}\ket{1} \nonumber
\end{align}

The problem is that these expressions are only equivalent when either $\alpha=0$ or $\beta=0$, which means the machine only works for $\ket{\psi} = \ket{0}$ or $\ket{\psi} = \ket{1}$. A quantum operation that clones the computational basis states $\ket{0}$ and $\ket{1}$ cannot also clone superpositions of those states. It follows that no quantum operation exists that can perfectly copy an arbitrary quantum state. We now argue that this places significant physical constraints on possible quantum agents.

\subsection{Initial challenges for quantum agency}

For the purposes of this paper, we have defined a \textit{purely quantum system} as one that evolves unitarily in a coherent regime without decoherence or collapse. We will assume here that the environment is also purely quantum, although later we will consider weakening that assumption. As explained in the introduction, we take these `purely quantum' systems to be characterised by the fact that there is no preferred basis, which is the regime in which the no-cloning theorem applies. While a set of  microscopic quantum systems in a coherent regime may well have a specific basis which plays a special role in the dynamics, this alone does not mean that generic systems from this set will typically have states drawn from this basis: in order to get convergence toward states which are diagonal in the preferred basis we require decoherence, which occurs either when the system itself is sufficiently large or when it is in regular interaction with a large external environment. Small coherent systems, by contrast, do not exhibit such convergence. We therefore focus on the regime without an emergent preferred basis, which we regard as a hallmark of classicality. 

The appeal to this regime allows us to exclude Everettian macroscopic agents from the discussion. In the Everett interpretation \citep{everett1957relative}, there is no collapse, and closed physical systems always evolve unitarily. In the modern form of the theory \citep{wallace2012emergent}, agents, as well as the worlds they exist within, emerge as a result of decoherence. Within these worlds, decoherence effectively selects a preferred basis, typically the coarse-grained position basis, such that everything is prepared in that basis and can be copied in that basis. Agency is therefore possible in the Everettian context in the decoherent regime, even if it is not possible for purely quantum systems. Our results can therefore be understood as constraining what it is possible for quantum agents to achieve in the case where no preferred basis is singled out.

Agency as we have defined it involves two different stages of copying. First the agent must copy information from the environment, to create or to update a world-model. Then the agent must create copies of that model for the purpose of deliberation. The first stage already leads to problems for a quantum agent, since the agent cannot simply copy information from an environment populated by unknown quantum states. It is possible to envision a quantum agent that makes imperfect copies of information in its environment, and we will come back to this case when we discuss approximate cloning strategies, but for now let us suppose that the agent aims to have a perfect copy of the environment state. In a quantum context, the only way to achieve this for a generic environment state is not to copy it at all, but rather to \textit{remove} it wholesale from the environment: for example, the agent might perform a swap operation to move the state $| \phi \rangle$ of some environment system into its memory, as in the ``swap sensation" described in \citet{lupu2024qubits}. 
 
But this leads to a further problem: unless there is some prior reason to expect that this state is the same as, or similar to,  the state of some other system in the environment, there will simply be no point in using the state $| \phi \rangle $ to decide how to act, because it is no longer in the environment and thus modelling it will give no useful information about the consequences of acting on the environment.\footnote{In addition to the swap sensation, \citet{lupu2024qubits} also consider von Neumann measurement without the collapse step, where the observer becomes entangled with the system, and the observer's information about the system is captured by the correlations between the observer and the system. In some sense, the observer has a record of the system. However, this record cannot then be used in deliberation or action in a purely quantum system, for the kinds of reasons we are about to give, and so would not amount to a world-model in our sense.} So for now let us restrict our attention to putative quantum agents existing in environments which contain many copies of the same or similar states, so that an agent can take in one of these systems and use it to predict the effects of operating on the correlated environment system. Of course, one might think this is a somewhat unrealistic case which cannot be expected to occur in the real world, but our point here is to show that even under this unrealistic idealization it is still impossible to have a meaningful quantum agent. 

Once the agent has taken an environmental state into itself, it will also need to create copies of that state in order to model the impact of its actions. And if it wants to also keep records to be used in future modelling, even more copies will be needed. For example, suppose the system has taken in information from the environment in the form of a quantum state $| \psi \rangle$, which is the same as, or similar to, the states of other systems in the environment. Suppose that there are two actions, $A_0$ and $A_1$, that the system can take. One might at first imagine creating two copies of  $| \psi \rangle$ and acting on each of them with unitary operations designed to represent the consequences of performing the actions $A_0, A_1$ respectively: $| \psi \rangle \rightarrow U_0 | \psi \rangle = | \psi_x \rangle$, $| \psi \rangle \rightarrow U_1 | \psi \rangle = | \psi \rangle_y$. Then the agent can compare the results to some metric which represents the desirability of the outcome - for example, perhaps the representation is designed such that more desirable outcomes are closer to the $z$ axis - and thus decide which action to perform on the systems in the environment. 

However, clearly this approach would require cloning, so it is not possible. One might think that instead of literally copying the state we could perform a superposition of the two operations $A_0$ and $A_1$ on a single copy of the state, and then use the resulting superposed state to determine which action produces a better outcome - but in fact this will not work. For example, if we act on the state $| \psi \rangle$ with an equal superposition of the two operations $U_0, U_1$ (which can still be a unitary operation under certain special conditions) then the result will be $\frac{1}{\sqrt{2}} ( U_0 |\psi \rangle + U_1 | \psi \rangle  ) =  \frac{1}{\sqrt{2}} (  |\psi_x \rangle +  | \psi_y \rangle  )$. But then the problem is that  $\frac{1}{\sqrt{2}} (  |\psi_x \rangle +  | \psi_y \rangle  )$ is the same as  $\frac{1}{\sqrt{2}} (  |\psi_y \rangle +  | \psi_x \rangle  )$, so once this operation has been performed there is no way to discern which of the two states $|\psi_x \rangle$  and $ | \psi_y \rangle$ in this representation comes from $U_0$ and which comes from $U_1$ unless the agent already knows in advance that $|\psi_x \rangle$  comes from $U_0$ and $y$ comes from $U_1$, and the agent presumably doesn't know that before performing the operation, since otherwise there would be no point in doing the modeling at all.

One might think that we could instead use a non-equal superposition of the two operations in order to make the possibilities distinguishable, resulting in something like $ \frac{1}{\sqrt{3}} U_0 |\psi \rangle + \sqrt{\frac{2}{3}} U_1 | \psi \rangle   =  \frac{1}{\sqrt{3}}  |\psi_x \rangle + \sqrt{\frac{2}{3}}  | \psi_y \rangle$. But note that the coefficients will look different in different bases, so unless we already know what basis we are supposed to be looking in, we still won't be able to extract useful information from this state. 
 
So the no-cloning theorem necessarily pushes the possibilities for quantum agency in one of two directions. We might consider quantum agents with merely approximate world-models, or we might consider quantum agents existing in very special environments consisting of a considerable number of copies of the same quantum state, such that the agent can in effect produce `copies' just by taking more states from the environment.  Of course, one might think this is an even more unrealistic case, but again, our point is to show that even under this highly unrealistic idealization it is still impossible to have a meaningful quantum agent.  We first consider the latter, and then the former.

\subsection{Quantum agents with unlimited perfect copies \label{copies}}

In the case where the agent has access to unlimited perfect copies, it can satisfy our first two conditions for agency. However, the third agency condition is still a problem. In this section we will demonstrate that there is no quantum operation which reliably selects and implements the `best' action, even when the agent has successfully performed modelling to establish what that action is.

We assume the agent's environment consists of many separable copies of the same unknown quantum state $\ket{\psi}$. The agent extracts two such copies and applies different unitary operations $U_1$ and $U_2$ to each:

\begin{equation}
    \ket{\psi} \otimes \ket{\psi} \xrightarrow{U_1 \otimes U_2} U_1\ket{\psi} \otimes U_2\ket{\psi}.
\end{equation}

This represents the agent preparing the input for evaluating the performance of actions $U_1$ and $U_2$ on an environmental state. Each unitary represents a candidate action, and its performance is evaluated by how close the resulting state is to a preferred target state, $\ket{0}$. 

The agent is defined this way because part of our notion of agency is that the agent must deliberate in order to achieve its goals - that is, it has a limited set of actions available to it and none of them is guaranteed to reliably achieve the goal in all circumstances, so it must take in information from the world to decide which one of the actions will best achieve the desired outcome. Thus our concept of a quantum agent involves a system which aims to prepare a certain fixed state  $\ket{0}$ and which has only a specific set of unitary operations available that it can perform on its environment. In particular, a system which has the capability to use adiabatic computing to reliably reset a system in an unknown state back to the ground state would not be an agent in this sense, since it would never have any need for deliberation.

  Let us express the joint output state in the computational basis:

\begin{equation}\label{joint}
    U_1\ket{\psi} \otimes U_2\ket{\psi} = (a\ket{0}  + b\ket{1}) \otimes (c\ket{0} + d\ket{1})
\end{equation}
where the amplitudes $a, b, c, d$ depend on $\ket{\psi}$ and the chosen unitaries.

The first qubit corresponds to the result of applying $U_1$, and the second to the result of applying $U_2$. For example, if $U_1\ket{\psi} = \ket{0}$ and $U_2\ket{\psi} = \ket{1}$, then $a=d=1$ and $b=c=0$ so the resulting state is $\ket{01}$, indicating that $U_1$ succeeded and $U_2$ failed. 

The agent will perform a controlled unitary $C_U$ that uses the two-qubit state $ac\ket{00} + ad\ket{01} + bc\ket{10} + bd\ket{11}$ as control. A third copy of $\ket{\psi}$ serves as the target. In principle we would like this operation to perform either $U_1$ or $U_2$ on the target, depending on which of them has produced an output closer to $|0\rangle$.  However, it is straightforward to show that there is no quantum operation which will achieve this for arbitrary input states. 

If there were such an operation, then it would have to act as follows:

\begin{itemize}

 \item If control = $\ket{01}$, apply $U_1$ to the target. Thus $\ket{01}\otimes\ket{\psi} \rightarrow \ket{01} \otimes U_1 \ket{\psi}$. 
    \item If control = $\ket{10}$, apply $U_2$ to the target. Thus $ \ket{10}\otimes\ket{\psi} \rightarrow \ket{10} \otimes U_2 \ket{\psi}$. 

    \item For now suppose we do not care what the operation does when the control is $\ket{00}$ or $\ket{11}$, so we will simply say $\ket{00}\otimes\ket{\psi} \rightarrow \ket{?}$ and $\ket{11}\otimes\ket{\psi} \rightarrow \ket{?'}$.
    \item If control =     $ (a\ket{0}  + b\ket{1}) \otimes (c\ket{0} + d\ket{1})$ where $|a| \geq |c|$, apply $U_1$ to the target, and if $|a|<|c|$, apply $U_2$ instead. Thus $ (a\ket{0}  + b\ket{1}) \otimes (c\ket{0} + d\ket{1}) \rightarrow \ket{z} \otimes U_1\ket{\psi}$ or $\ket{z} \otimes U_2\ket{\psi}$, where $\ket{z}$ is any state of the two control qubits. 

    \end{itemize}

But it can be seen that an operation satisfying the first two conditions will contradict the last.  From linearity, this operation would have to act as: $(a\ket{0}  + b\ket{1}) \otimes (c\ket{0} + d\ket{1}) \rightarrow ac \ket{?} + ad \ket{01} \otimes U_1 \ket{\psi}  + bc \ket{10} \otimes U_2 \ket{\psi} + bd \ket{?'}$, and as we now show, this state cannot generally be of the form $\ket{z} \otimes U_1\ket{\psi}$ or $\ket{z} \otimes U_2\ket{\psi}$.  

Let us set $|u_1\rangle \equiv U_1 \ket{\psi} =  a\ket{0} + b\ket{1}$ and write $\ket{u_2} \equiv U_2\ket{\psi} = c\ket{0} + d\ket{1}$.  If we consider the $|a|\geq|c|$ case, in order for 
\begin{equation}
    ac\ket{?} + ad\ket{01}\ket{u_1} + bc\ket{10}\ket{u_2} + bd\ket{?'} = \ket{z}\ket{u_1}
\end{equation}
to be true,
we must have
\begin{equation}
    ac\ket{?} + bc\ket{10}\ket{u_2} + bd\ket{?'} = \ket{z'}\ket{u_1}.
\end{equation}
The controlled unitary must map the four orthogonal control states to four orthogonal states, so $\ket{01}\ket{u_1}$, $\ket{10}\ket{u_2}$, $\ket{?}$, and $\ket{?'}$ are mutually orthogonal.  The question is, can we choose $\ket{?}$ and $\ket{?'}$ so that the above expression is satisfied?

To deal with the $\ket{10}\ket{u_2}$ term, we will need $\ket{?} = \ket{10}\ket{u_{2}^\bot}$, such that $ac\ket{?} + bc\ket{10}\ket{u_2} = c\ket{10}(a\ket{u_2} + b \ket{u_{2}^\bot}) = c\ket{10}\ket{u_1}$.  This works if $c=0$, or if  $a\ket{u_2} + b \ket{u_{2}^\bot} = \ket{u_1} = a\ket{0} + b\ket{1}$, meaning $\ket{u_2} = \ket{0}$ and $\ket{u_2^\bot} = \ket{1}$, and thus $c=1$.  So, for general values of $c$, the expression is not satisfied.

Alternatively, we could deal with the $\ket{10}\ket{u_2}$ using $\ket{?'} = \ket{10}\ket{u_{2}^\bot}$, such that $bd\ket{?'} + bc\ket{10}\ket{u_2} = b\ket{10}(c\ket{u_2} + d \ket{u_{2}^\bot} )= b\ket{10}\ket{u_1}$.  This works if $b=0$ or if $c\ket{u_2} + d \ket{u_{2}^\bot} = \ket{u_1} = a\ket{0} + b\ket{1} = (c^2 + |d|^2)\ket{0} + d(c-c^*)\ket{1}$, meaning $a = c^2 + |d|^2 = 1$ and $b = d(c-c^*) = 0$ (noting that we can choose $c$ to be real by fixing a global phase).  So, for general values of $b$, the expression is not satisfied.

Thus it is impossible for the quantum agent to implement an operation which always chooses and applies the best unitary to the target. If the operation has the correct action for the basis states $\ket{01}$ and $ \ket{10}$, then for cases where the controls end up in superpositions of $\ket{0}$ and $\ket{1}$, some mixture of $U_1$ and $U_2$ will be applied to the target. Moreover, in general the target qubit will end up entangled with the control qubits, so we will not end up with a pure state of the target qubit at all. 

Here we construct an operation of this kind which has the correct action on the basis elements. To act on a third copy of $\ket{\psi}$ based on the deliberation state in (\ref{joint}), the action of $C_U$ is defined by:

\begin{itemize}
    \item If control = $\ket{01}$, apply $U_1$ to the target.
    \item If control = $\ket{10}$, apply $U_2$ to the target.
    \item If control = $\ket{00}$ or $\ket{11}$, apply some unitary (examples considered below).
\end{itemize}

In general, we define a \textit{quantum agency circuit} to be a specific unitary operation that acts on $N$ copies of an environment state as $Q_A = C_U (U_1 \otimes...\otimes U_{N-1})$, which `deliberates' on the first $N-1$ by testing $U_i$ and then using them as controls in a corresponding $C_{U}$, and which  attempts to change the last copy to the desired target state (e.g. $|0\rangle$). 

Consider a case where the control qubits do not favor a single unitary $U_i$.  One simple option is to arbitrarily choose one of the unitaries $U_i$ for such cases, but the arbitrariness makes this choice of $C_U$ seem undesirable.

Rather than arbitrarily choosing between them, we can perform a superposition of the operations in question.  In general, unitary operations cannot be superposed to give unitary operations, but there are some special cases where $\sum_i \alpha_i U_i$ is unitary, for particular sets of complex coefficients $\alpha_i$ and unitaries $U_i$.  We consider such a case here.

Applying $\alpha_1 U_1 + \alpha_2 U_2$ in both the $\ket{00}$ and $\ket{11}$ cases, with $|\alpha_1|=|\alpha_2|$, reflects a kind of symmetry in the agent's assessment: in the $\ket{00}$ case, both $U_1$ and $U_2$ appear to bring $\ket{\psi}$ close to the desired outcome $\ket{0}$, and so are equally good; in the $\ket{11}$ case, both appear equally poor. In both cases, the agent lacks a decisive reason to favor one action, and thus applies an equal superposition of the two unitaries.

This leads to the combined three-qubit output state:

\begin{align}
    \ket{\Psi}_{\text{out}} =
    &\, ac\ket{00} \otimes (\alpha_1 U_1 + \alpha_2 U_2)\ket{\psi} \\
    &+ ad\ket{01} \otimes U_1\ket{\psi} \\
    &+ bc\ket{10} \otimes U_2\ket{\psi} \\
    &+ bd\ket{11} \otimes (\alpha_1 U_1 + \alpha_2 U_2)\ket{\psi}.
\end{align}

Note that in general, this state may or may not be entangled between the control and target qubits, depending on the values of $a$, $b$, $c$, and $d$. For instance, if $a=d = 1$ and $b=c=0$, the state is separable: the agent determinately applies $U_1$ to the target. But for general values of the coefficients, the system usually becomes entangled.

To illustrate the action of the controlled unitary, we now consider a special case where:

\begin{equation}
    U_1\ket{\psi} \otimes U_2\ket{\psi} = \Big(\sqrt{\tfrac{9}{10}}\, \ket{0} + \sqrt{\tfrac{1}{10}}\, \ket{1}\Big) \otimes \ket{1}  =\sqrt{\tfrac{9}{10}}\, \ket{01} + \sqrt{\tfrac{1}{10}}\, \ket{11}.
\end{equation}

This means that, with high squared amplitude, the agent's assessment favors $U_1$ over $U_2$, but there is some amplitude associated with both unitaries being judged poor.

Feeding this control state into the controlled operation $C_U$ acting on a third copy of $\ket{\psi}$ yields:

\begin{equation}
    \ket{\Psi}_{\text{out}} =
    \sqrt{\tfrac{9}{10}}\, \ket{01} \otimes U_1\ket{\psi} + \sqrt{\tfrac{1}{10}}\, \ket{11} \otimes (\alpha_1 U_1 + \alpha_2 U_2)\ket{\psi}.
\end{equation}

This state represents the agent’s action being guided by a superposition of evaluations: with high squared amplitude, it acts on the target qubit using $U_1$, but with smaller squared amplitude, it acts using an equal mixture of $U_1$ and $U_2$.

To see where this leaves the target qubit, we compute its reduced density matrix by tracing out the two control qubits. This gives us a mixed state that describes what operation the agent effectively applied to the third copy of $\ket{\psi}$.

To obtain the reduced density matrix of the target qubit (the third register), we trace out the first two qubits:

\begin{align}
    \rho_{\text{target}} &= \Tr_{12} \left( \ket{\Phi}\bra{\Phi} \right) \\
    &= \tfrac{9}{10} \, U_1\ket{\psi}\bra{\psi}U_1^\dagger
    + \tfrac{1}{10} (\alpha_1 U_1 + \alpha_2 U_2)\ket{\psi}  \bra{\psi}(\alpha_1^* U_1^\dagger + \alpha_2^* U_2^\dagger)  \nonumber
\end{align}

This reduced state describes the actual effect of the agent’s control procedure on the environment. With 90\% probability, the agent applied $U_1$ to $\ket{\psi}$. With 10\% probability, it applied the superposition $\alpha_1 U_1 + \alpha_2 U_2$. 

To consider a concrete case, suppose the agent tests the identity operation $I$ against the Pauli-$X$ gate, which flips the basis states $\ket{0} \leftrightarrow \ket{1}$. We begin with an environment state $\ket{\psi} = \ket{1}$. In this scenario, $U_1 = I$ fails to bring the state closer to the target state $\ket{0}$, while $U_2 = X$ succeeds exactly. That is:

\begin{align}
I\ket{1} &= \ket{1}, \\
X\ket{1} &= \ket{0}.
\end{align}

The agent prepares the deliberation state:

\begin{equation}
\ket{\psi}_{\text{delib}} = I\ket{1} \otimes X\ket{1} = \ket{1} \otimes \ket{0} = \ket{10}.
\end{equation}

Interpreting the left qubit as the outcome of $U_1$ and the right as the outcome of $U_2$, this state signals that $U_2$ succeeded and $U_1$ failed. According to the control scheme, when the deliberation register is in the state $\ket{10}$, the agent applies $U_2 = X$ to a fresh copy of $\ket{\psi}$. Since $X\ket{1} = \ket{0}$, the result is exactly the target state. The total three-qubit state becomes:

\begin{equation}
\ket{\Psi}_{\text{out}} = \ket{10} \otimes X\ket{1} = \ket{10} \otimes \ket{0}.
\end{equation}

This is a product state (i.e., not entangled), and tracing out the control qubits yields the pure state $\ket{0}$ for the target qubit. The fidelity\footnote{Fidelity is a standard measure of how close a quantum state $\rho$ is to an intended target state $\sigma$, given by the most general expression: $F(\sigma, \rho) = \big(\textrm{tr}\sqrt{\sqrt{\rho} \sigma \sqrt{\rho}} \big)^2$. It expresses the probability that one state will pass a test to identify as the other.} with the target state is thus:

\begin{equation}
F(\ket{0}) = |\langle 0 |0 \rangle |^2 = 1.
\end{equation}

This demonstrates that when the deliberation state yields a decisive verdict — namely, when one unitary produces the target state and the others produce orthogonal states — the agent can act with certainty of success.

This circuit also gives $F=1$ if the input state is $|0\rangle$, and these two results together show what we consider to be \textit{the classical limit of quantum agency}, where the agent knows which basis the environment states will belong to, and can thus design a $Q$ that achieves $F=1$ for all basis states.  We call this the classical limit because all classical observables commute, and have a single joint eigenbasis, and knowing this basis enables the creation of arbitrarily many clones of an environment state to use for deliberation.  This is also the case where $Q$ succeeds in choosing and enacting the single best unitary, and thus fully satisfies all three agency conditions.  This $Q$ also does not entangle the target qubit with the control qubits.  This contrasts with the case above, where the deliberation produces nonzero coefficients for $|00\rangle$ or $|11\rangle$, and the agent does not choose a single unitary, but rather enacts a mixture of unitaries and entangles the control qubits with the target qubit.  Even if this gets close to the desired target state, it does not meet our necessary condition, because the agent did not choose the single best option.

This definition of the classical limit of $Q$ also generalizes to systems whose Hilbert spaces have arbitrary dimension.  For example, for qutrits, if the classical basis is known, then it is always possible to choose three unitaries such that deliberation on any basis state results in outputs $\ket{0}$, $\ket{1}$, and $\ket{2}$ (in some order), so there is always exactly one $\ket{0}$ in the control sequence.  For qu$d$its, the $d$ deliberation unitaries likewise give output states $\ket{0},\ket{1},\ldots,\ket{d}$ for any classical basis state, so there is always a single $\ket{0}$ in the control sequence.

Returning to qubits, we now consider a case where the environment state is:

\begin{equation}
\ket{\psi} = \sqrt{\tfrac{3}{4}} \ket{0} + \sqrt{\tfrac{1}{4}} \ket{1}.
\end{equation}

Applying $I$ and $X$ gives:

\begin{align}
I\ket{\psi} &= \sqrt{\tfrac{3}{4}} \ket{0} + \sqrt{\tfrac{1}{4}} \ket{1}, \\
X\ket{\psi} &= \sqrt{\tfrac{3}{4}} \ket{1} + \sqrt{\tfrac{1}{4}} \ket{0}.
\end{align}

The resulting deliberation state is:

\begin{align}
\ket{\psi}_{\text{delib}} &= \left(\sqrt{\tfrac{3}{4}} \ket{0} + \sqrt{\tfrac{1}{4}} \ket{1} \right) \otimes \left(\sqrt{\tfrac{3}{4}} \ket{1} + \sqrt{\tfrac{1}{4}} \ket{0} \right) \\
&= \tfrac{\sqrt{3}}{4} \ket{00} + \tfrac{3}{4} \ket{01} + \tfrac{1}{4} \ket{10} + \tfrac{\sqrt{3}}{4} \ket{11}.
\end{align}

When the deliberation register is in $\ket{01}$ or $\ket{10}$, the agent applies a single unitary (either $U_1$ or $U_2$) to the target state. When the deliberation register is in $\ket{00}$ or $\ket{11}$, it applies a unitary superposition of unitaries $\tfrac{1}{\sqrt{2}}(I \pm iX)$. Applying the controlled operation to the target qubit results in the following entangled three-qubit state:

\begin{align}
\ket{\Psi}_{\text{out}} &= \tfrac{\sqrt{3}}{4} \ket{00} \otimes \tfrac{1}{\sqrt{2}}(I + iX)\ket{\psi} \\
&\quad + \tfrac{3}{4} \ket{01} \otimes I \ket{\psi} \\
&\quad + \tfrac{1}{4} \ket{10} \otimes X \ket{\psi} \\
&\quad + \tfrac{\sqrt{3}}{4} \ket{11} \otimes \tfrac{1}{\sqrt{2}}(I - iX)\ket{\psi}.
\end{align}

Tracing out the control qubits yields a mixed state for the target. In this case, the fidelity of the resulting state with respect to the desired target state $\ket{0}$ is approximately:

\begin{equation}
F(\ket{0}) \approx 0.625.
\end{equation}

Even though one of the unitaries, $I$, is a relatively good match for $\ket{\psi}$ (bringing it mostly toward $\ket{0}$), and $X$ is somewhat worse, the entanglement of control and target introduces a kind of incoherence that ``corrupts'' the target. This degradation in performance arises from the agent's inability to extract a clean classical verdict from its quantum deliberation.

We should also note that there is some freedom in choosing the coefficients $\alpha_i$ for unitary superpositions, so there are still many subtle variations possible for these circuits.

\subsection{Approximate cloning strategies}

We have argued that the prohibition on cloning quantum states is a central obstacle standing in the way of the possibility of quantum agents. But of course, although \emph{perfect} cloning of quantum states is impossible, there exist various techniques which allow some approximation of cloning. These include: 

\begin{itemize}
\item State-dependent cloning: it is possible to construct a unitary operation which will perfectly clone any member of a set of orthogonal states. For more general sets of states it is possible to construct state-dependent cloning operations which will optimally clone that particular set of states, although perfect cloning is impossible for non-orthogonal sets \citep{PhysRevA.57.2368}. So if we are provided with a guarantee that the state in question will belong to a certain set, then we can sometimes clone more effectively.

\item Probabilisitic-exact cloning: it is possible to construct a unitary operation together with a measurement which with some probability $p$ succeeds in cloning an input state from a set of linearly independent states \citep{Duan_1998}. It can also be arranged that if the cloning succeeds, then a reference system $S$ is put into the state $|0\rangle$, so if we allow post-selection then we can clone with perfect accuracy (at the cost of failing some number of attempts and losing the input state).

\item Deterministic-imperfect cloning: it is possible to construct a unitary operation which takes a qubit in any state $|\psi \rangle$ and produces two systems with density matrix $\rho = F | \psi \rangle \langle \psi | + (1 - F) |\psi^{\perp}  \rangle \langle \psi^{\perp}  | $ with fidelity $F = 5/6$ \citep{Bu_ek_1996,Scarani_2005}.

\end{itemize}

The possibility of state-dependent cloning suggests an interesting variation on the scenario we described in section \ref{copies}. There, we considered an agent in an environment made up of a number of systems which all have the same unknown state, and we saw that this leads to numerous difficulties. Alternatively,  we could consider  an agent in an environment made up of systems with states which are guaranteed to belong to some orthogonal basis, known in advance to the agent. In that case, the system can have a built-in unitary which perfectly clones states belonging to that basis, and thus it can take in an environment state, produce as many copies as needed, and proceed with modelling the effect of various unitary operations as we have described. Moreover, if we assume that the `target state' belongs to this orthogonal basis, and that the agent therefore only models unitaries which take elements of the basis to other elements of the basis, then it is possible to define an evaluation operation which will always deterministically select and implement the best operation in a way which results in a pure state for the target qubit, as in the first example of section \ref{copies}. So in this special case, the agent can perform modelling and select appropriate actions with perfect accuracy, making this ideal territory for agency. This is not surprising since classical agents are in effect something like quantum agents existing within an environment made of systems belonging to some orthogonal basis, i.e. the basis favoured by decoherence (usually the coarse-grained position basis), and indeed in the Everett interpretation and other unitary-only interpretations it is presumably the case that classical agents literally are just quantum agents existing in a decoherence regime which ensures that all relevant systems are in well-defined states in the coarse-grained position basis. So from this example we can understand how classical agency arises out of the quantum world in appropriate circumstances. 

The possibility of probabilistic-exact cloning relies on being able to make a classical measurement, which the quantum agent is not able to do. Of course, we can still perform the probabilistic cloning procedure and simply treat the measurement as a von Neumann measurement modelled unitarily with the outcome recorded in some register, but then this simply becomes a version of deterministic-imperfect cloning and thus it will be subject to the same fidelity restrictions. Note also that probabilistic cloning can be achieved only for a set of linearly independent input states, so it would in any case not suffice for a quantum agent whose environment is completely unknown. 

For the possibility of deterministic-imperfect cloning, consider first the case of  symmetric universal $N \rightarrow M$ cloning, i.e. a unitary transformation which takes $N$ copies of the input state and produces $M$ copies, with the same fidelity for all of the copies and for any input state.  The protocol described above with fidelity 5/6 is optimal for the  $1 \rightarrow 2$ case. In the next section we present some numerical results for the case where the agent uses $1 \rightarrow M$ cloning to produce copies to use for world-modelling. 

More generally for the $N \rightarrow M$ case, the maximum fidelity is $\frac{MN + M + N}{M(N + 2)}$ \citep{Gisin_1997}, so given one original the fidelity of each of the $M$ copies to the original is $\frac{2M + 1}{3M}$.  If we only have one copy of an unknown environment state to work with, but we still want to use a unitary quantum agency circuit to get it into a desired state, we may be able to succeed in making enough symmetric clones.  These clones retain the correct angle on the Bloch sphere, but the length of the Bloch vector is reduced.

However, it can be seen that the fidelity of each clone to the unknown environment state rapidly decreases and asymptotes toward $2/3$ as we create more clones.
Note that  $1/2$ is the average fidelity we get if we simply draw a random pure state from the Bloch sphere  and use that random state as our `clone,' using the Haar measure \citep{Scarani_2005}.
Moreover, we can achieve average fidelity of $3/4$ in the $M = 2$ case if we just choose a state at random to use as the `copy,' because in that case we are averaging over fidelity $1$ (the original) and fidelity $1/2$ (the random `copy'). So we can see that the optimal cloning procedure with fidelity 5/6 for $M = 2$ is not doing very much better than simply choosing a state at random. And the asymptotic average fidelity of 2/3 achievable by deterministic-imperfect cloning is not much better than the result that we get from guessing at random.

Thus, deterministic-imperfect cloning does not allow the quantum agent to adequately satisfy our second agency condition. For as the number of possible actions the agents wishes to deliberate over increases, the fidelity of the necessary set of clones gets worse.

 Note also that in general we must use an ancilla to implement deterministic-imperfect cloning - for example, the optimal operation for $1 \rightarrow 2$ cloning also uses one qubit as an ancilla (and the ancilla is left in the optimal anti-clone, i.e. the best estimation of the result of performing a NOT gate on the input qubit). This is in addition to the reference qubit on which the copy will be prepared. So in order to repeatedly implement cloning of this kind the quantum agent will need to have a pool of reference systems and also a pool of ancilla systems, with all of the references prepared in some fixed state and all of the ancillae prepared in some fixed state.  It is questionable that a purely unitary agent would have access to such a pool, but we will grant it for the purpose of this analysis.

It is also possible to implement asymmetric universal cloning, where the cloning is equally good for any input state but some of the copies have better fidelity than others. For two copies with fidelities $F_A$ and $F_B$, these fidelities must satisfy the inequality $\sqrt{(1 - F_A)(1 - F_B)} \leq \frac{1}{2} - (1 - F_A) - (1 - F_B)$ \citep{Scarani_2005,PhysRevLett.81.5003}. So our quantum agent could potentially choose an approach where the first copy has better fidelity and subsequent copies become progressively worse. However unless the agent has some prior knowledge about which operations are more likely to be successful, it is unlikely that this would be a more successful approach, since performing an operation which in fact does better at the task on a worse copy could result in incorrectly concluding that this operation performs worse than one which has been implemented on a better copy.

\subsection{Some quantum agency circuits and their performance}

\begin{figure}[ht!]
    \centering
    \begin{tabular}{|l|ccc|ccc|}
    \hline
    Scenario&&$N$ copies&&&$N$ clones&\\
    \hline\hline
        $Q_{(I)X}$, $N=2$ & worst & average & best & worst & average & best\\
        \hline
        Fidelity & $1/2$ & $2/3$ & 1 &  2/3  & 2/3 &  2/3\\
        Bloch Vector Length & 0 & 0.73399 & 1&  1/3  &0.49421   &0.74536\\
        Bloch Angle Error (rad) & $\pi/2$ & 0.97095  & 0 &  1.10715 &  0.64282  &                 0\\
            \hline\hline
        $Q_{IX}$, $N=3$ & worst & average & best & worst & average & best\\
        \hline
        Fidelity & $1/2$ & $2/3$ & 1&  2/3  & 2/3 &  2/3\\
        Bloch Vector Length & 0 & 0.73399 & 1&  1/3 &   0.45326   & 0.64788\\
        Bloch Angle Error (rad) & $\pi/2$ & 0.97095  & 0 &   1.03048  & 0.58182 &                 0 \\
    \hline\hline
        $Q_{IHX}$, $N=4$ & worst & average & best & worst & average & best\\
        \hline
        Fidelity & $0.43562$ & $0.69372$ & $0.92678$ & 0.62644   &0.69394  & 0.76144
\\
        Bloch Vector Length & 0.30530 &0.69391 & 0.95040 &   0.35600  & 0.46061  & 0.58729
\\
        Bloch Angle Error (rad) & 1.86017 & 0.96929  & 0.11297 &  1.02859  & 0.51454  & 0.12499
 \\
\hline\hline
        $Q_{IX'Y'Z'}$, $N=5$ & worst & average & best & worst & average & best\\
        \hline
        Fidelity & 0.44774 &   0.65882 &  0.76955 &                 0.62840 &  0.65802   &0.68765
 \\
        Bloch Vector Length & 0.18519  & 0.51126   & 0.66668 & 0.25952   &0.32018   &0.37996\\
        Bloch Angle Error (rad) & 1.76471 & 0.82512   &   0 &    0.26909  &0.14712 & 0.04660
\\
        \hline
    \end{tabular}
    \caption{This table attempts to characterize the performance of four different quantum agency circuits, by evaluating our 26 test inputs.  These circuits use between $N=2$ and $N=5$ environment states, as shown.  The left block shows the case where the agent is given $N$ perfect copies of the pure state $|\psi\rangle$, and the right block shows the case where the agent must make $N$ symmetric clones of a single $|\psi\rangle$, which really means creating a symmetric entangled state of $N$ qubits and acting the deliberation unitaries on the first $N-1$ of this state. $Q_{IX}$ deliberates using $I$ and Pauli $X$ for its two control qubits.  $Q_{IHX}$ deliberates using $I$, Hadamard $H$, and $X$ for its three control qubits. $Q_{IX'Y'Z'}$ deliberates using $I$, and rotated Paulis $X'$, $Y'$ and $Z'$ for its four control qubits.  Lastly, $Q_{(I)X}$ deliberates using only $X$ for its single control, and blindly applies $I$ for control $|1\rangle$.  Because $Q_{(I)X}$ does not deliberate on at least two choices, we think of it as a proto-agent, but it is noteworthy that it performs just as well as $Q_{IX}$ with perfect copies, and very similarly for clones.}
    \label{fig:Qperf}
\end{figure}

\begin{figure}[ht!]
    \centering

\begin{tabular}{cc}
     \begin{tabular}{c}
       $Q_{(I)X}$, $N=2$ \\
        \begin{tabular}{|cc|}
        \hline
            Unitary: & $C_U(X \otimes I)$ \\
            \hline
             Controls & Applied Unitary\\
             \hline
             $|0\rangle$ & $X$ \\
            $|1\rangle$ & $I$\\
             \hline
        \end{tabular}  \\\\
        $Q_{IX}$, $N=3$ \\
        \begin{tabular}{|cc|}
        \hline
            Unitary: & $C_U(I \otimes X \otimes I)$ \\
            \hline
             Controls & Applied Unitary\\
             \hline
             $|00\rangle$ & $(I + iX)/\sqrt{2}$ \\
            $|01\rangle$ & $I$\\
            $|10\rangle$ & $X$\\
            $|11\rangle$ & $(I - iX)/\sqrt{2}$\\
             \hline
        \end{tabular} \\\\

            $Q_{IHX}$, $N=4$ \\
        \begin{tabular}{|cc|}
        \hline
            Unitary: & $C_U(I \otimes H \otimes X \otimes I)$ \\
            \hline
             Controls & Applied Unitary\\
             \hline
             $|000\rangle$ & $(I +iH +iX)/\sqrt{3+\sqrt{2}}$ \\
            $|001\rangle$ & $(I-iH)/\sqrt{2}$\\
            $|010\rangle$ & $(I+iX)/\sqrt{2}$\\
            $|011\rangle$ & $I$\\
            $|100\rangle$ & $(H +X)/\sqrt{2+\sqrt{2}}$ \\
            $|101\rangle$ & $H$\\
            $|110\rangle$ & $X$\\
            $|111\rangle$ & $(I +iH -iX)/\sqrt{3-\sqrt{2}}$\\
             \hline
        \end{tabular} \\
    \end{tabular}  &

     \begin{tabular}{c}

            $Q_{IX'Y'Z'}$, $N=5$ \\
        \begin{tabular}{|cc|}
        \hline
            Unitary: & $C_U(I \otimes X' \otimes Y' \otimes Z' \otimes I)$ \\
            \hline
             Controls & Applied Unitary\\
             \hline
             $|0000\rangle$ & $(I +iX'+iY'+iZ')/2$ \\
            $|0001\rangle$ & $(I +iX'+iY')/\sqrt{3}$\\
            $|0010\rangle$ & $(I -iX'+iZ')/\sqrt{3}$\\
            $|0011\rangle$ & $(I +iX')/\sqrt{2}$\\
            $|0100\rangle$ & $(I -iY'-iZ')/\sqrt{3}$ \\
            $|0101\rangle$ & $(I +iY')/\sqrt{2}$\\
            $|0110\rangle$ & $(I +iZ')/\sqrt{2}$\\
            $|0111\rangle$ & $I$\\
              $|1000\rangle$ & $(X'+Y'+Z')/\sqrt{3}$ \\
            $|1001\rangle$ & $(X'-Y')/\sqrt{2}$\\
            $|1010\rangle$ & $(Z'-X')/\sqrt{2}$\\
            $|1011\rangle$ & $X'$\\
            $|1100\rangle$ & $(Y'-Z')/\sqrt{2}$ \\
            $|1101\rangle$ & $Y'$\\
            $|1110\rangle$ & $Z'$\\
            $|1111\rangle$ & $(I -iX'-iY'-iZ')/2$\\
             \hline
        \end{tabular} \\
    \end{tabular}
\end{tabular}
    \caption{For each example quantum agency circuit, the complete unitary is given, along with the details of the controlled unitary $C_U$.  The target state is $|0\rangle$, so where a deliberation unitary produces a $|0\rangle$, $C_U$ applies that unitary, or a superposition of all deliberation unitaries that gave $|0\rangle$.  $X'$, $Y'$, and $Z'$ are the Pauli matrices rotated to have equal expectation values $1/\sqrt{3}$ for the target state $|0\rangle$.}
    \label{QA}
\end{figure}

As we have seen, there are many approaches to constructing optimal quantum agency circuits, and the task of exploring all possibilities is quite sophisticated.  Here we choose a handful of examples or circuits $Q_A$ which are intended to be somewhat representative, and we evaluate their performance.  To do this, we consider 26 environment states, whose Bloch vectors are permutations of $(\pm 1, 0, 0)$, $(\pm 1, \pm 1, 0)/\sqrt{2}$, and $(\pm 1, \pm 1, \pm 1)/\sqrt{3}$, which give a reasonably symmetric covering of the Bloch sphere.  

For the simplest case, we provide $N$ copies of the pure environment state $|\psi\rangle$, and evaluate the circuit for all 26 states.  From this we take the worst, best, and mean values of the fidelity, the Bloch length, and the Bloch angular error, relative to the target state $|0\rangle$, and report these to characterize the performance of $Q_A$ in Fig. \ref{fig:Qperf}.  The detailed form of the agency circuits $Q_A$ we tested are given in Fig. \ref{QA}.

Remarkably, for each of our four test circuits $Q_A$, the average fidelity is roughly the same for the case of $N$ pure copies as it is for the case of the $N$-qubit entangled state generated by the cloning machine from a single copy.
Even though the reduced density matrix of each qubit within this $N$-qubit entangled state has lowered fidelity, when we feed in this state, $Q_A$ uses the correlations to circumvent this extra deterioration.  It is important to emphasize that $Q_A$ is not actually deliberating on $N-1$ separate world-models here, because the qubits housing the world-models are entangled.

The average fidelity of the deliberation circuit in our test cases is around 2/3, and this is only slightly better than just taking a random state from the environment, which gives an average fidelity of 1/2. We think it may be possible to engineer a $Q_A$ that performs better, but there are many possibilities, so this is an open problem.

So, $Q_A$ can do better than 50/50 on getting the desired output state, and this may be adequate for some decision processes, but we don't think it rises to the level of agency, wherein one deliberates on separate world-models and chooses the single best action.  However, as discussed above, if the agent knows the basis that the environment states all belong to, then they can design a circuit $Q_A$ that meets our necessary conditions for agency, and makes the best choice with fidelity 1.

As a final thought here, we think these $Q_A$ are an interesting set of circuits which may be worth examining for other applications.  The data in Figure \ref{fig:Qperf} show the result of tracing out all of the control qubits to get a reduced density matrix for the target qubit, but as with the cloning machine, if we don't trace out all other qubits, there may be less fidelity and purity loss, and perhaps all qubits can be fed into whatever process was intended to use the target state.  In this sense, the fidelity of $Q_A$ may actually be better than it appears.

\section{Discussion}

We have formulated and motivated three necessary conditions on agency, and have shown that they cannot be adequately satisfied all at once in a purely quantum system. Our analysis has revealed that the emergence of a preferred basis is a crucial classical ingredient required for the emergence of agency. As we now show, these results challenge quantum theories of agency, consciousness, and free will, forcing such proposals to identify explicitly where the necessary classical resources enter into their account, and to explain why the quantum resources they invoke are not thereby rendered redundant. 

We begin with models of quantum \textit{agents}. According to \citet{lupu2024qubits}, there is ``no \textit{a priori} reason to rule out fully quantum agents with coherent quantum memories." The authors then introduce an entirely quantum notion of measurement or observation, called a \textit{sensation}, to ``account for quantum agents that experience the world through quantum sensors". After quantifying the information gain and disturbance of a sensation, it is shown that sensations always disturb at least as much as they inform (see also \citet{pang2024information}). This is an interesting result and is consistent with ours because it concerns observation or sensing, not decision making. However, the authors also describe their results as simply the ``first steps in studying fully quantum agents" and that ``the processes by which quantum agents might decide and act using those observations is still ripe for exploration". Since, as we have argued, `deciding' requires cloning, our results show that this may not be possible. While the environment and sensations may be quantum, subsequent processing may need to be classical, as in \citet{kewming2021designing}. In the terminology of \citet{dunjko2016quantum}, where the system is labeled CC, CQ, QC or QQ, depending on whether the agent (first symbol) or the environment (second symbol) are classical (C) or quantum (Q), our results challenge the possibility of QQ. Our results also challenge CQ since the classical agent cannot clone arbitrary quantum states in the environment. They may also challenge QC insofar as the quantum agent does not know the preferred basis which all the environmental states belong to. Indeed, a small quantum system may not have access to a reference frame aligned with the classical basis in which the environment is prepared (\cite{bartlett2003classical,gour2008resource}), and in that case the environment will in effect be in an unknown quantum state relative to it, so from the quantum agent's point of view QC might end up looking the same as QQ. QC is therefore an unstable category: if Q knows the basis, then QC is not meaningfully different from CC; if Q does not know the basis, then QC is not meaningfully different from QQ. 

We turn now to quantum theories of \textit{consciousness}. These have been motivated in a variety of ways. The ``hard problem'' of consciousness (\cite{chalmers1995facing}) has led many to think that a classical description of the brain is insufficient to explain consciousness and that a quantum ingredient may be necessary. \cite{penrose1994shadows} famously used Godel's theorem to argue that conscious understanding must depend on non-computable brain processes, which Penrose connected to wave function collapse. Some have claimed that the binding problem (how seamlessly integrated a conscious experience is) is best explained by entanglement (\cite{neven2024testing, georgiev2024causal, faggin2024irreducible}). Some have claimed that the subjective privacy of one's conscious experience might be explained by the experience being a pure quantum state and therefore not cloneable (\cite{d2022hard, Georgiev2025-GEOQIT-3}). Some have proposed quantum physical correlates of consciousness in the brain, such as nuclear spins (\cite{fisher2015quantum, swift2018posner}), or microtubules (\cite{hameroff2014consciousness}). The feasibility of these proposals is sometimes defended by pointing  to alleged cases of ``quantum biology", such as energy transfer in photosynthesis (\cite{fassioli2014photosynthetic}), bird navigation by earth's magnetic field (\cite{hiscock2016quantum}), tunneling in biochemical reactions (\cite{klinman2013hydrogen}), and superradiance in tryptophan (\cite{babcock2024ultraviolet})).

Although our results pertain to agency, there are three important connections to consciousness. First, agents typically base world models on what they experience. An agent experiences their environment and then creates copies of environmental information to weigh the consequences of different possible actions. This copying process would be compromised if the experience is an uncloneable quantum state. Second, some have argued that agency is constitutive of consciousness (\cite{liu2024agency}), suggesting that if the former cannot be purely quantum, then neither can the latter. Finally, even if agency is not constitutive of consciousness, there is agential phenomenology i.e. what it feels like to deliberate and choose, and this will be a challenge to explain in purely quantum terms. If the deliberation necessarily takes place in a classical regime while consciousness is associated with the coherent quantum regime, there is a prima facie problem about how we can have conscious experiences of agency.

According to \citet{d2022hard} and \citet{Georgiev2025-GEOQIT-3}, one's current state of consciousness (which might be an experience of the environment) is an uncloneable pure quantum state. A motivation given for this view is that the unclonability explains why our experiences feel subjectively private, or not fully knowable to others. But then, how could one copy the information in the experience to memory or for deliberation? The answer in \citet{d2022hard} is that the quantum experience state is (somehow) duplicated en masse, inaccessibly to the conscious agent. Then, the agent's brain performs tomography on all (or many of) these copies, to derive an approximate copy to be stored in classical memory. Then, presumably, classical deliberation can take place. As they say ``It is also likely that the conscious act of memorising an experience may be achieved by actually repreparing the ontic state multiple times in the quantum buffer, and performing the infocomplete test multiple times, thus with the possibility of memorising a (possibly partial) tomography of the state'' (p167). This need for many copies of the experience (in a ``quantum buffer") is similar to our agent surrounded by many copies of 
$\psi$. The difference is that while our agent is a ``pure" quantum agent, this agent invokes tomography and so measurement-induced collapses. Furthermore, our case was contrived: we know our reality doesn't consist of many copies of the same thing. But then, where in their picture do all the copies arise from?  If experience is really a pure unknown quantum state, then it is surely impossible for the brain to produce a large number of copies of it: that would violate no-cloning. And if experience is some kind of quantum state which is always prepared in a fixed basis, then it is not really meaningfully `quantum' any more (e.g. we could no longer appeal to unclonability to explain why experiences are subjectively private). 

The most well-known quantum theory of consciousness is the Orch OR (“orchestrated objective reduction”) model of Hameroff and Penrose (\citet{hameroff2014consciousness,hameroff2022orch,penrose2022new}). Orch OR attributes consciousness to orchestrated quantum state reductions occurring within microtubules. In this view, tubulin proteins within microtubules can exist in coherent superpositions of alternative conformations, forming extended entangled states across many neurons. These states are claimed to be shielded from environmental decoherence long enough for large-scale quantum computation to occur. When a particular objective threshold (set by Penrose's quantum-gravitational criterion) is reached, the superposition undergoes a non-computable ``objective reduction" (OR) event, which is identified with a discrete moment of conscious experience. The ``orchestration" refers to the microtubular quantum computations that bias or tune the pre-collapse superposition, shaping the conscious content. 

Orch OR does not provide an explicit account of deliberation or agency. One natural interpretation is that the theory is of the QC type, where the agent is fully quantum (deliberation included) and the environment is classical. Classical information enters the brain (from an effectively collapsed environment) and then enters a quantum circuit (implemented across many microtubules): the quantum agent. The quantum agent needs a cloneable world-model. Either the quantum agent knows what basis the world-model information is given in, or it doesn't.
If it doesn't, then our results refute Orch OR: it would make the world-model cloning required for agency impossible. If it does know the basis, then there is a further dilemma. Either the quantum circuit implements something like our controlled unitary, which knows the basis in the sense that all action unitaries transform preferred basis states into preferred basis states, or it does something else. If it implements something like our controlled unitary, then we have the classical limit of the quantum system, which falls into the purely classical CC category, and the quantum computations seem redundant. Alternatively, the quantum circuit may implement something else to ensure that it really does create superpositions and entanglement, making it recognizably quantum. But then Orch OR must explain how a completely coherent quantum agent is going to have enough internal resources to store information about a classical basis, which is non-trivial (\cite{bartlett2003classical,gour2008resource}).

Alternatively, Orch OR could move away from a QC model and adopt a HC model where H means the agent is a quantum classical hybrid: agency is classical while consciousness is quantum. But then we need an account of how these two ``modules" or circuits interact. The view may face the same problems as the previous views that treated conscious experiences as pure states: how can their information be cloned for agency to take place? The collapsed microtubule state that resulted from prior quantum computations is insufficient. The same post-collapse state could be associated with different experiences. For the experience is the noncomputable collapse event itself, not the subsequent computable post-collapse state. It is unclear how these noncomputable collapse events could contribute information to deliberation and memory. 

The integrated information theory (\cite{albantakis2023integrated}), or IIT for short, is a leading neuroscientific theory of consciousness that is the focus of ongoing experimental tests (\cite{cogitate2025adversarial}). The theory is classically formulated, and allows one to calculate the integrated information $\Phi$ of certain idealized classical neural networks, a supposed measure of their consciousness level. \citet{albantakis2023computing} extend IIT into the quantum domain by developing a new quantum measure of integrated information (specifically, a measure of ``mechanism" integrated information, which is a component of the $\Phi$ measure). (For related approaches, see \cite{zanardi2018towards,kleiner2021mathematical}.) The goal is to develop a measure of the quantum integrated information, and so, consciousness, of a quantum state. Quantum IIT raises the same question as other quantum consciousness proposals: can a system whose conscious contents are instantiated in a quantum state use that very state to compare alternative futures or evaluate possible actions? Doing so would require creating clones of the state to support deliberation, in conflict with the no-cloning theorem. The form of IIT's $\Phi$ measure is motivated and constrained by a set of assumptions or ``axioms" about consciousness. If IIT were to include a new axiom, stating that \textit{conscious states must be cloneable}, then it might avoid the need to consider quantum integrated information altogether.

Finally, we consider theories of \textit{free will}. Many have wondered how we could be ``free'' in a deterministic physical universe. Incompatibilists hold that free will and determinism are mutually exclusive and, consequently, that we act freely only if determinism is false. Libertarian accounts of free will insist that we do sometimes act freely, and so develop indeterministic models of free action. Quantum indeterminacy and wavefunction collapse often play a role in these accounts. Although our results only concern ``purely" quantum agents that do not collapse, our results still place constraints on how deliberation can occur in these models. We illustrate with the most widely discussed libertarian theory of free will, that of Robert \cite{kane1985free,kane1998significance, kane2007libertarianism, kane2019complex}. According to Kane, an agent's free will depends on the agent being able to perform so-called self-forming actions. Such actions involve a kind of motivational conflict, where one is motivated to perform action A as well as some incompatible action B. For example, when making a moral choice, one might be motivated both to do what is right and to do what is in one's self-interest, and may thereby struggle with the deliberation. It is in this motivational conflict that Kane postulates that quantum randomness enters to resolve the conflict, leading the agent towards an undetermined action. Unfortunately, the actual physical mechanisms by which randomness enters are left quite vague. To avoid cloning issues, one might postulate that it happens quite late in the deliberation process. Perhaps first, one copies models of the relevant situation, and calculates the consequences of different actions, and only then do quantum effects enter. In that case, the copying process may have been completed and the resulting indecision somehow creates a superposition of intending to do A and intending to do B, which may then randomly collapse to one of these options. However, Kane does not advocate such a simplistic model. In part to avoid certain objections (relating to the ``problem of luck"), Kane endows the deliberation process with many ongoing collapses that build up and lead to an undetermined action. In particular, he postulates that one's ongoing motivational effort to choose one option over an alternative desirable option creates random nudges toward one option.\footnote{``The choice one way or the other is undetermined because the process preceding and potentially terminating in it (i.e., the effort of will to overcome temptation) is indeterminate" (Kane, 1996, 128).} It is unclear how this process could be physically described. If the copying takes place in a classical regime the resulting intentions will also be in a classical regime. Furthermore, one often creates new world models during deliberation: I might be undecided between doing A vs B, which might cause me to consider doing C, requiring a new world model. But how can I consider C if I've entered a superposition of intending to do A and intending to do B? We leave this is as open challenge to such accounts.

\section*{Acknowledgments}
 Authors are listed in alphabetical order and contributed equally. We would like to thank Ian Durham for extensive discussions, and Justin Dressel, Matt Leifer, Aaron Schurger, and Uri Maoz for helpful feedback.  This project/publication was made possible through the support of Grant 63209 from the John Templeton Foundation. The opinions expressed in this publication are those of the authors and do not necessarily reflect the views of the John Templeton Foundation.

  \bibliographystyle{apalike}
 \bibliography{references}{}

\end{document}